# Magnetization reversal in bent nanofibers of different cross-sections

Tomasz Blachowicz[1], Andrea Ehrmann[2]

[1] *Silesian University of Technology, Institute of Physics – Center for Science and Education, 44-100 Gliwice, Poland*

[2] *Bielefeld University of Applied Sciences, Faculty of Engineering and Mathematics, 33619 Bielefeld, Germany*

Artificial ferromagnetic nanofiber networks with new electronic, magnetic, mechanical and other physical properties can be prepared by electrospinning and may be regarded as the base of bio-inspired cognitive computing units. For this purpose, it is necessary to examine all relevant physical parameters of such nanofiber networks. Due to the more or less random arrangement of the nanofibers and the possibility of gaining bent nanofibers in this production process, elementary single nanofibers with varying bending radii, from straight fibers to those bent along half-circles, were investigated by micromagnetic simulations, using different angles with respect to the external magnetic field. As expected from the high aspect ratios and the resulting strong shape anisotropy, all magnetization reversal processes took place via domain wall processes. Changing the cross-section from circular to a circle-segment or a rectangle significantly altered the coercive fields and its dependence on the bending radius, especially for the magnetic field oriented perpendicular (90°) to the fiber axes. In all three cross-sections, an angle of 45° between the fiber orientation and the external magnetic field resulted in the smallest influence of the bending radius. The shapes of the longitudinal and transverse hysteresis curves showed strong differences, depending on cross-section, bending radius and orientation to the magnetic field, often depicting distinct transverse magnetization peaks perpendicular to the fibers for fibers which were not completely oriented parallel to the magnetic field. Varying these parameters thus provides a broad spectrum of magnetization reversal processes in magnetic nanofibers and correspondingly scenarios for a variety of fiber-based information processing.

# I. Introduction

Similar to quantum computers which are expected to solve a defined class of mathematical problems on very short time-scales,[1-3] a novel adaptive computation technology can be estimated to calculate different problems similar to the human brain and thus in a much more flexible and efficient way than common technology.[4-6] Besides developing hard- and software architecture further, however, only few attempts have been made to create completely new approaches for a new generation of computers.

One approach – also being implemented in soft- rather than in hardware until now – is the "cognitive computing", as postulated by many research centers,[7,8] i.e. creating computers which may work similarly to the human brain, with far lower energy consumption than recent computers and more effectively by combining "thinking" (i.e. the processor) and "memorizing" (i.e. data storage), as it is done in the human brain. This approach necessitates completely new hardware architectures, with adaptive / active materials which can fit their physical / chemical properties, e.g. electrical connections or their general arrangement, to recent necessities. Early works of Heinz von Foerster and W. R. Ashby about biology-based computing and self-organizational systems include a broad spectrum of ideas which can be used here.[9-12] In their pioneering works they defined what are the natural and fundamental principles of self-organizing systems with internal intelligence and logics resulting from interaction between vast number of constituting elements.

One possibility to create such a new "cognitive" computer hardware is based on a large number of highly interconnected simple processors, connected with large amounts of low-latency memory, giving thus up the classical von Neumann architecture with the strict separation of memory and processor.[13] For example, Grollier *et al.* propose networks of connected memristors to realize bioinspired or "neuromorphic" computing.[14] Another possibility is based on creating bio-inspired nanofiber networks of fibers prepared in a form of mats, partly mimicking the brain's geometry and its data storage, processing and conducting properties. Such a nanofiber system firstly necessitates understanding the physical properties of the single fibers and afterwards merging them to a network, creating new connections and thus new physical properties.

Storing data in magnetic nanofibers is not a new idea. Such magnetic nanofibers with different cross-sections can be created in diverse ways, e.g. by Focused-ion-beam (FIB) milling,[15,16] e-beam lithography[17] or even guided self-assembly of block copolymer thin films.[18] Structuring such nanowires with diminutions of the fiber radius can be used to create "nanotraps", i.e. positions at which domain walls are pinned and depinned again for certain magnetic field strengths and orientations. In this way, it could be possible to create hysteresis loops with multiple well-defined switching transitions.[19] Magnetic nanowires can also be used to create logic gates or to "clone" domain walls[20] as well as to store information in domain walls for memory applications.[21,22] For these applications, domain walls are usually moved by an electrical current which adds a correlation between magnetic and electronic systems that may be useful in magnetic memory applications,[23-25] e.g. in the form of the Racetrack memory.[26-28]

In purely magnetic fields, nanowires show different magnetization reversal modes, depending on their aspect ratios and geometric arrangement with respect to other nanowires, partly giving rise to unusual effects such as steps in the hysteresis loops and additional stable magnetic states at remanence which can be used, e.g., for magnetic storage applications.[29-35]

While straight nanofibers and magnetic rings have been investigated in diverse research groups, examinations of bent nanofibers, however, are scarce. Recently, Garg et al. have shown that the curvature of nanowires strongly influences domain wall propagation driven by currents.[36] Domain wall oscillations at constant magnetic field in curved magnetic nanowires with circular cross-section were shown by Moreno et al.[37] For magnetic ribbons of helicoid and Möbius shape, Gaididei et al. found geometry-induced magnetic states.[38] An effective magnetic anisotropy and effective Dzyaloshinskii-like interactions were calculated for arbitrary curved magnetic wires.[39]

While these studies concentrate on special shapes and cross-sections, our investigation aims at understanding the influence of the curvature and cross-section of the elementary component of the fiber-based network on magnetic properties, magnetization dynamics, spatial transfer of domain walls, occurrence and stability level of exotic magnetization states, all being possible elements of future information processing. For this, three different cross-sections (circular, circle-segment, and rectangular) were examined in combination with three orientations of the fibers with respect to the external magnetic field and several curvatures, starting from a straight line to a half-circle. This elementary approach reflects a number of effects in a vast set of fiber-based magnetic system.

## II. Experimental

The samples under examination in this article were designed using finite elements methods enabling realistic design of curvature. All fibers have a length of $l = 1570$ nm. The following cross-sections were taken into account:

a) circular cross-section – with the diameter $d = 60$ nm,
b) circle-segment – with the height equaling $h = 10$ nm and the base width $b = 60$ nm,
c) rectangular cross-section – with the width $w = 60$ nm and the height $h = 10$ nm.

The curvatures are defined here by the bending deflection $y_b$ which is depicted in Fig. 1. This value is varied between $y_b = 0$ nm (straight line) and $y_b = 500$ nm (semi-circle).

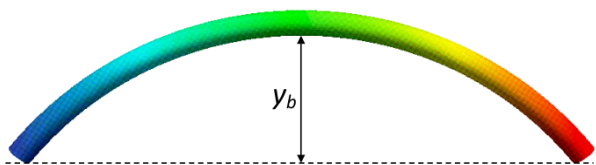

Fig. 1. Definition of the bending deflection $y_b$ in the middle of the fiber.

Simulations were performed with the LLG micromagnetic MagPar solver,[40] dynamically integrating the equation of motion. The simulation parameters were: material Co, damping coefficient 0.01, anisotropy constant $K_1 = 5.3 \cdot 10^5$ J/m$^3$, exchange constant $A = 1.03 \cdot 10^{-11}$ J/m, field sweep rate 10 kA/(m·ns), saturation polarization $J_s = 1.76$ T,[41,42] and the external field was swept in the range between ± 400 kA/m.

Sample orientation angles with respect to the external magnetic field are given with respect to the line through both ends of the fiber, i.e. the dotted line in Fig. 1. The hysteresis loops depicted in Section III are separated into the longitudinal magnetization component $M_L$ and the transverse magnetization component $M_T$, as defined by the orientation of the external magnetic field. This means that for a magnetic field orientation of 90°, $M_L$ is oriented perpendicular to a straight fiber, while $M_T$ is parallel to it. The magnetization "snapshots", on the other hand, are always related to the fiber orientation, with the x-orientation being parallel to the line through both ends of the fiber, i.e. the dotted line in Fig. 1, and y also being oriented in the curvature plane.

### III. Results and Discussion

Fig. 2 depicts the bending deflection dependent coercive fields $H_C$ of the fibers with circular, circle-segment, and rectangular cross-sections.

For the circular fibers – which can be assumed to be mostly created in an electrospinning process –, the magnetic field orientation parallel to the fiber (0°) shows nearly no influence of the deflection on the coercive fields until nearly a half-circle is reached. Only after increasing $y_b$ from 425 nm to 450 nm, a jump in the line is visible which may be correlated with a change in the magnetization reversal process. For a 45° orientation with respect to the magnetic field, a relatively steady decrease of the coercive fields with increasing bending is visible. For the 90° orientation, however, a significant dependence of $H_C$ on the deflection is visible, starting at vanishing coercive field for the straight fiber to a maximum at a deflection of 175 nm and decreasing again afterwards.

These findings are completely different for the samples with circle-segment cross-sections. Such shapes can be created, e.g., by e-beam lithography or Focused Ion Milling (FIB). Here, a maximum of $H_C$ can be found for the field orientation parallel to the fibers (0°) instead, while both other orientations show a relatively smooth curve without noticeable jumps etc. For all deflections smaller than 500 nm, the coercive fields for the 45° orientation can be found between those for the 0° and 90° orientations which is different from the fibers with circular cross-sections where such a rule does not apply.

Finally, for the rectangular cross-section, the general trends of the curves as well as the absolute values are similar to those simulated for the circle-segment fibers. The coercive fields for the 45° orientation are again in most cases between those gained by the other magnetic field angles. Such a rectangular shape could be created in an ideal e-beam lithography or FIB milling process, without undesired underetching, shadowing or other typical problems occurring in nanostructure preparation.

Rectangular and circle-segment cross-sections are meant as the extrema of typical lithography processes, while in electrospinning mostly circular fibers are created. Apparently, different magnetic properties can be expected from circular nanofibers, compared to usual lithographically produced nanowires with non-circular cross-sections, potentially opening a new field of applications.

These differences in the magnetization reversal processes and hysteresis loops will be examined more in detail in the next sub-sections.

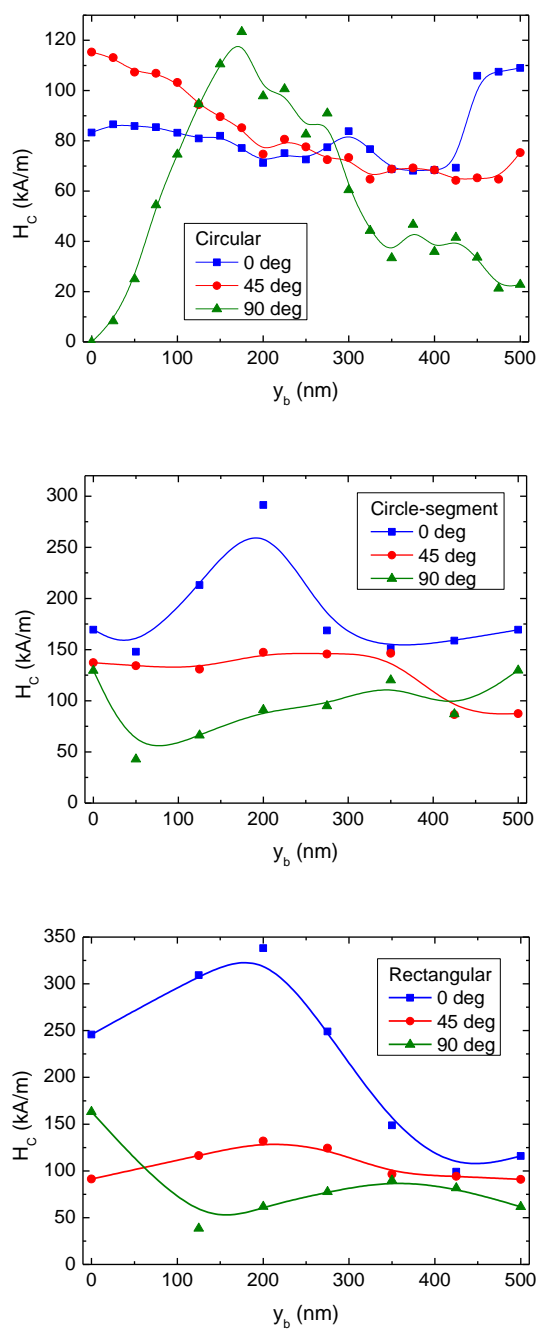

Fig. 2. (Color online) Coercive fields of fibers with different cross-sections and varying bending deflections, simulated for angles of 0°, 45°, and 90° with respect to the external magnetic field. Lines are only guides to the eye.

## A. Circular cross-section

Hysteresis loops for fibers with circular cross-section in a sample orientation of 0° are shown in Fig. 3. While a typical easy-axis loop is depicted for $y_b = 0$, steps in the loop become visible for $y_b = 425$ nm, vanishing again at the maximum deflection of 500 nm.

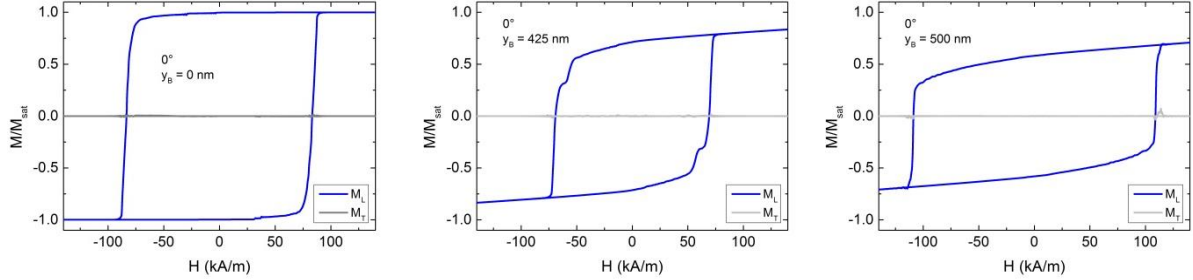

Fig. 3. (Color online) Longitudinal ($M_L$) and transverse ($M_T$) hysteresis loops of fibers with circular cross-section, simulated for an orientation of 0° with respect to the external magnetic field and different bending deflections $y_b$. $M_T$ is nearly constantly zero here.

The snapshots of the magnetization during the reversal process from positive to negative saturation show the reversal processes more in detail. As can be seen in Fig. 4 for $y_b = 0$, magnetization reversal starts at both ends of the fiber, with the domain walls moving to the middle of the sample and finally vanishing to complete magnetization reversal. It should be mentioned that the opposite magnetization orientations marked in the z magnetization orientation ("into" the page and "out of it") show a vortex-like structure surrounding the fiber axis, while the nearly vanishing marks in the y component show that this vortex is not completely symmetric. It should be mentioned that due to the rotational symmetry of this situation, this asymmetry is arbitrary; following the simulation back from negative to positive saturation revealed stronger evidence of a vortex structure in the y component of the magnetization.

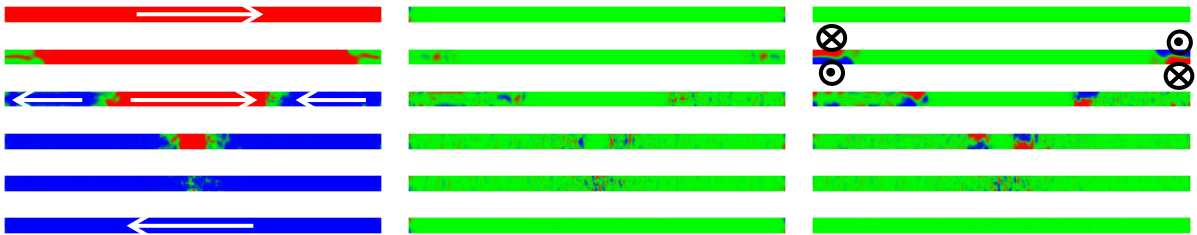

Fig. 4. (Color online) Snapshots of the magnetization for $y_b = 0$ nm and an angle of 0°, starting from positive saturation (top) to negative saturation (bottom), depicting x, y and z components of the magnetization (from left to right). Arrows are used to show magnetization reversal direction in some of the pictures for easier evaluation of these images.

For a bending deflection of $y_b = 425$ nm, magnetization reversal again starts at the ends of the fibers (Fig. 5). Now, a clear vortex structure is visible in the z component of the magnetization (marked with round arrows) which is dissolved after the first part of the magnetization reversal. This vortex state is much more extended than the vortex domain walls visible in Fig. 4. Finally, when both vortices get in contact, the domain wall between them vanishes, and magnetization reversal is completed.

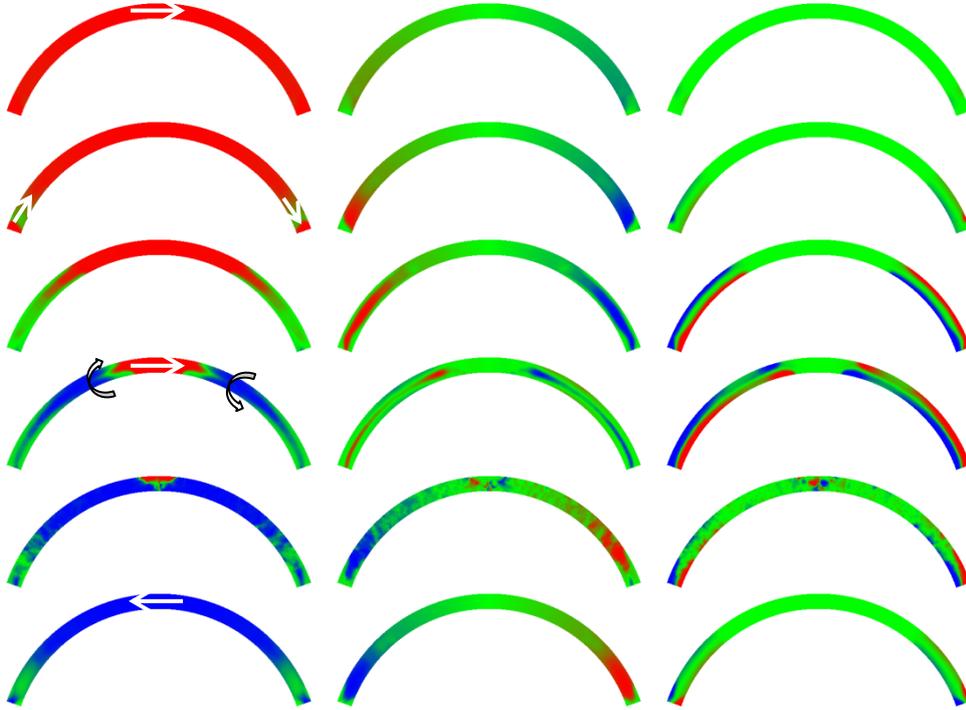

Fig. 5. (Color online) Snapshots of the magnetization for $y_b$ = 425 nm and an angle of 0°, starting from positive saturation (top) to negative saturation (bottom), depicting x, y and z components of the magnetization (from left to right).

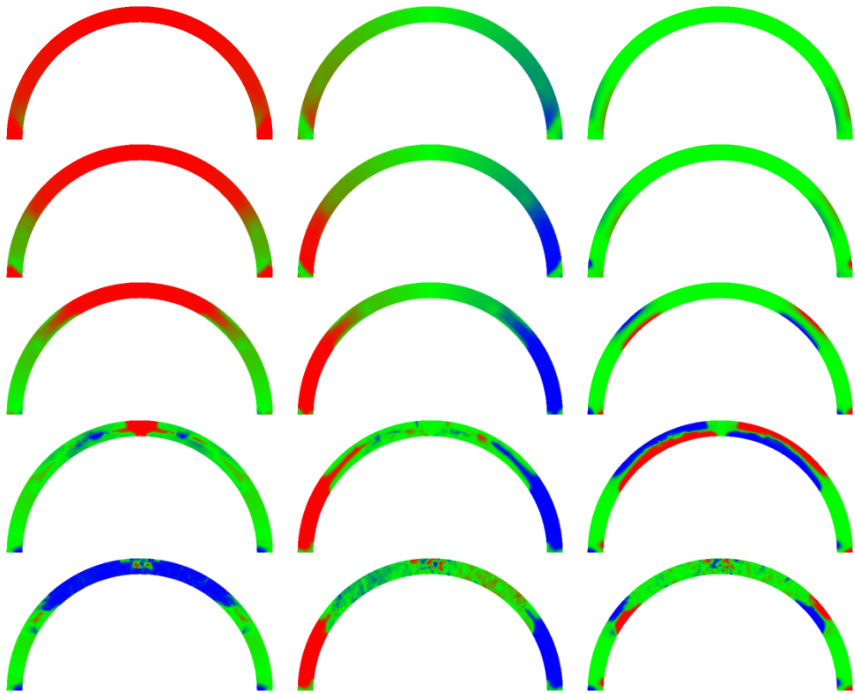

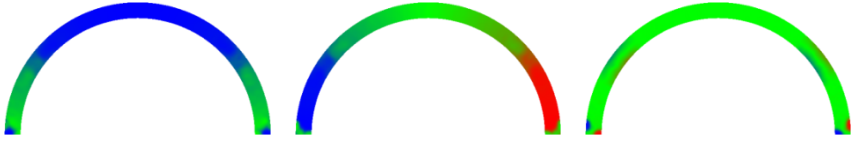

Fig. 6. (Color online) Snapshots of the magnetization for $y_b$ = 500 nm and an angle of 0°, starting from positive saturation (top) to negative saturation (bottom), depicting x, y and z components of the magnetization (from left to right).

For the largest bending deflection, nearly the same process occurs (Fig. 6). The only visible difference is the different orientation of the vortices which do not show a component along the external magnetic field here, opposite to Fig. 5 (cf. the fourth steps of both processes). Apparently the shape anisotropy blocks this vortex orientation for $y_b$ = 500 nm, thus impeding a faster magnetization reversal and increasing the coercive fields.

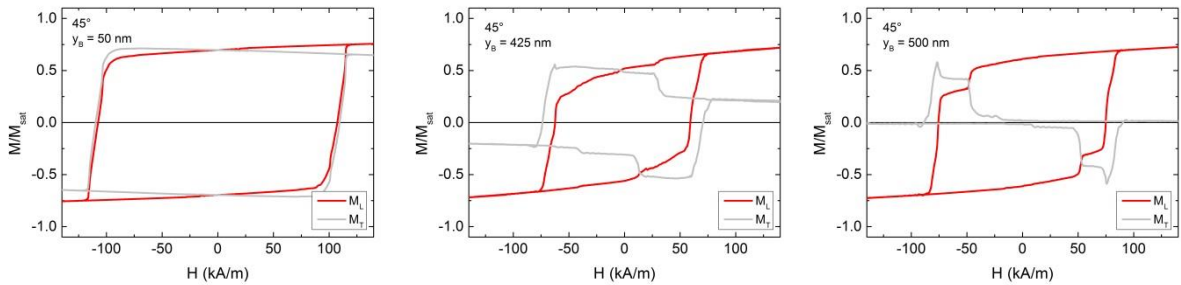

Fig. 7. (Color online) Longitudinal ($M_L$) and transverse ($M_T$) hysteresis loops of fibers with circular cross-section, simulated for an orientation of 45° with respect to the external magnetic field and different bending deflections $y_b$.

In the next simulation series, the angle with respect to the external magnetic field was set to 45°. Fig. 7 depicts the hysteresis loops simulated for different bending deflections. In the transverse loops it is clearly visible that for larger values of $y_b$, the transverse magnetization at saturation becomes smaller until it finally vanishes for the largest deflection.

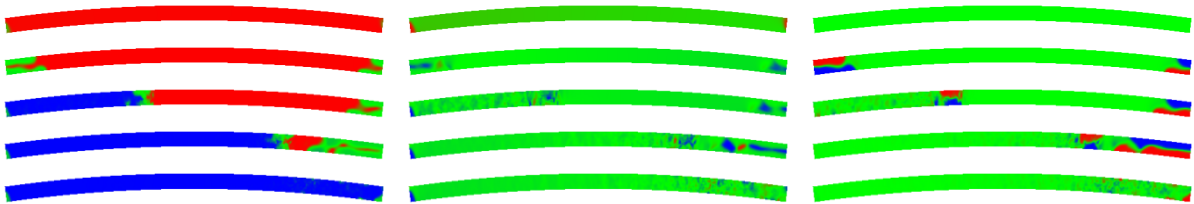

Fig. 8. (Color online) Snapshots of the magnetization for $y_b$ = 50 nm and an angle of 45°, starting from positive saturation (top) to negative saturation (bottom), depicting x, y and z components of the magnetization (from left to right).

For a relatively small deflection of 50 nm (Fig. 8), the magnetization reversal works similar to the process depicted in Fig. 4 for 0° and vanishing deflection. The asymmetry which is strongly visible here is introduced by the 45° angle with respect to the external magnetic field, resulting in one half of the fiber being oriented more parallel to the magnetic field and thus being reversed easier and the other half fiber being oriented differently.

A similar asymmetry is also visible in Fig. 9, depicting snapshots of the magnetization reversal of a fiber with 425 nm deflection. Going further to $y_b$ = 500 nm (Fig. 10), the vortex on the left side vanishes again, similar to the situation depicted in Fig. 6 for an angle of 0°.

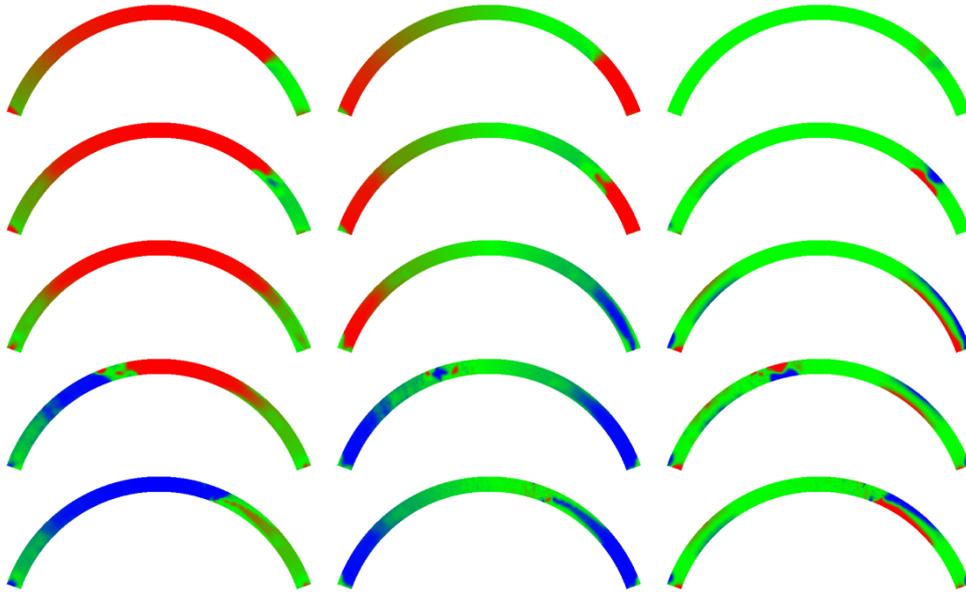

Fig. 9. (Color online) Snapshots of the magnetization for $y_b$ = 425 nm and an angle of 45°, starting from positive saturation (top) to negative saturation (bottom), depicting x, y and z components of the magnetization (from left to right).

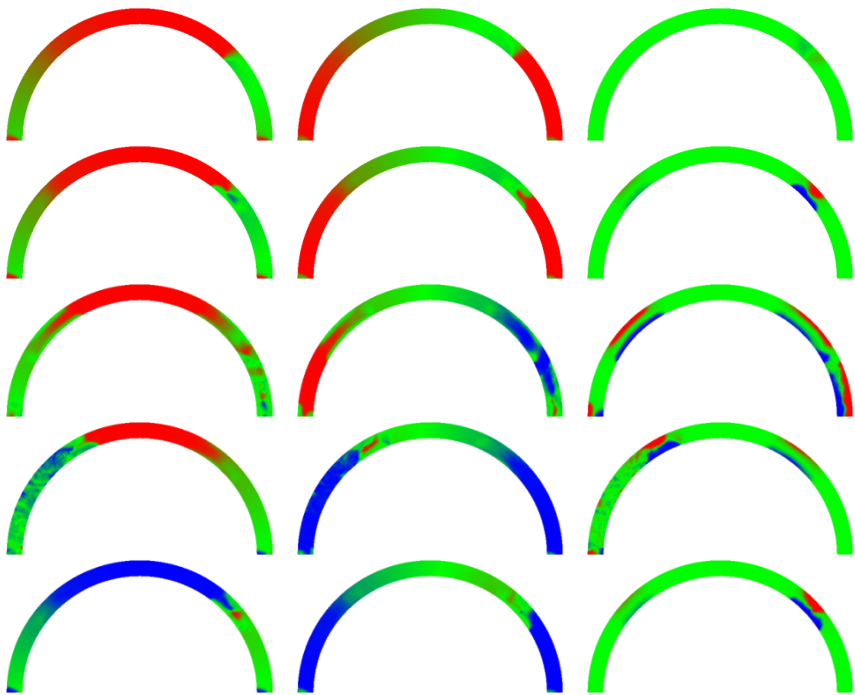

Fig. 10. (Color online) Snapshots of the magnetization for $y_b$ = 500 nm and an angle of 45°, starting from positive saturation (top) to negative saturation (bottom), depicting x, y and z components of the magnetization (from left to right).

Next, Fig. 11 depicts hysteresis loops for an angle of 90° with respect to the external magnetic field. Obviously, this direction is identical with the hard axis for $y_b$ = 0. For all bending deflections, large transverse peaks in different orientations are visible. It should be mentioned that the orientation of these peeks seems to be arbitrary; no system becomes visible by comparing the loops for all deflections under examination.

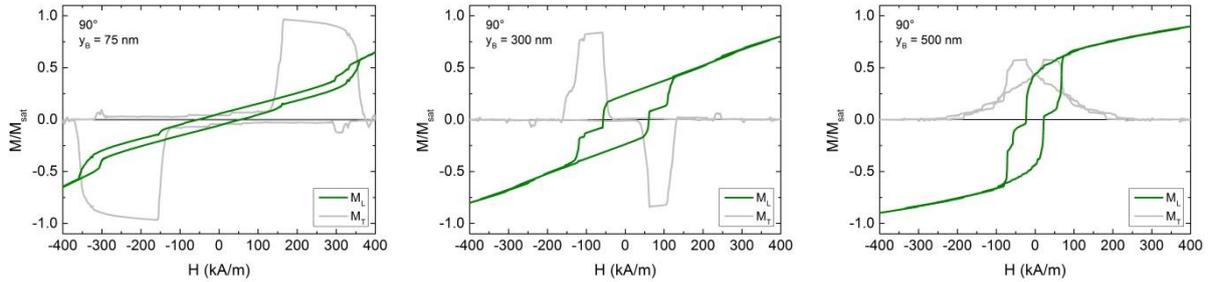

Fig. 11. (Color online) Longitudinal ($M_L$) and transverse ($M_T$) hysteresis loops of fibers with circular cross-section, simulated for an orientation of 90° with respect to the external magnetic field and different bending deflections $y_b$.

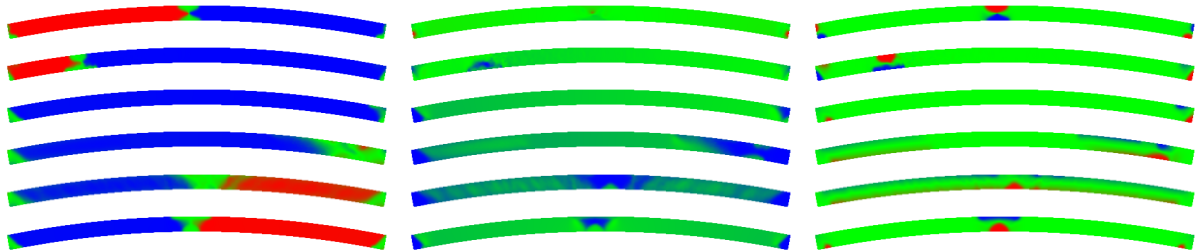

Fig. 12. (Color online) Snapshots of the magnetization for $y_b$ = 75 nm and an angle of 90°, starting from positive saturation (top) to negative saturation (bottom), depicting x, y and z components of the magnetization (from left to right).

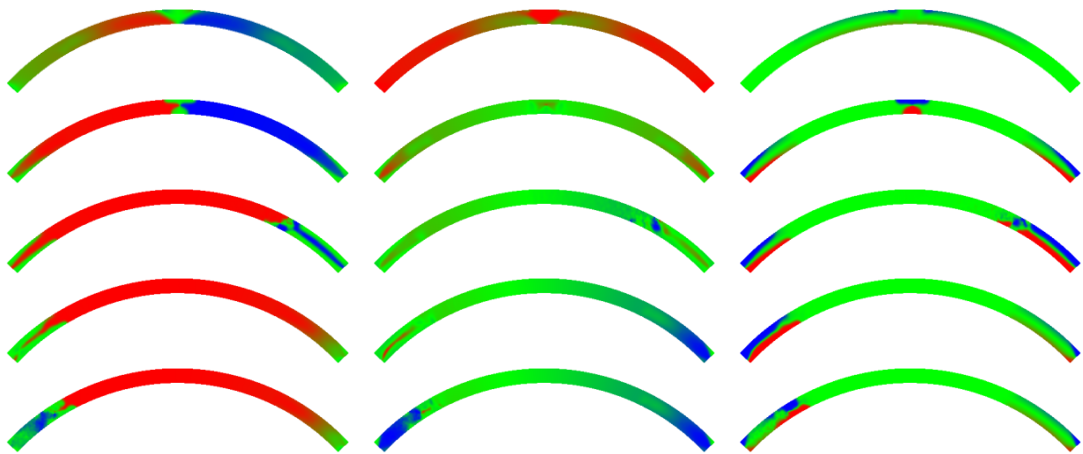

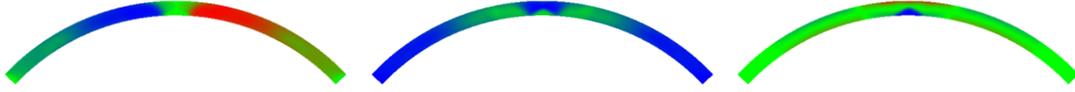

Fig. 13. (Color online) Snapshots of the magnetization for $y_b = 300$ nm and an angle of 90°, starting from positive saturation (top) to negative saturation (bottom), depicting x, y and z components of the magnetization (from left to right).

The corresponding snapshots of the magnetization reversal processes are depicted in Figs. 12-14. Depending on the deflection, more or less orientation into the y direction (i.e. the direction of the magnetic field) is visible. In the middle of the fiber there is always a domain wall during saturation, resulting from the necessity that the magnetization in both halves of the fiber is oriented towards y which is correlated with an orientation along +x for the left half and along –x for the right half of the fiber. Magnetization reversal happens by this domain wall moving to one end of the fiber, and another domain wall being moved to the middle of the fiber from the other end.

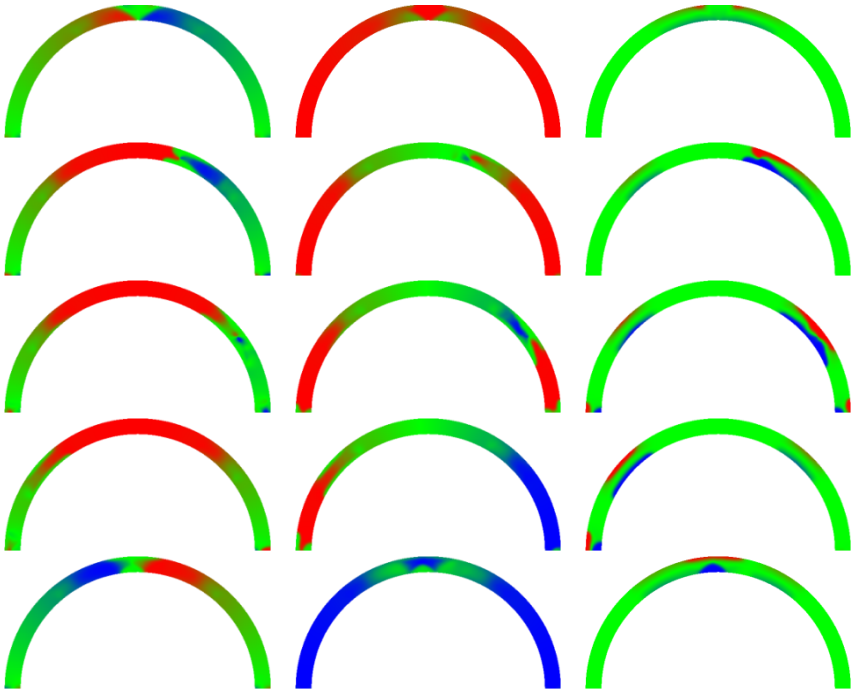

Fig. 14. (Color online) Snapshots of the magnetization for $y_b = 500$ nm and an angle of 90°, starting from positive saturation (top) to negative saturation (bottom), depicting x, y and z components of the magnetization (from left to right).

**B. Circle-segment cross-section**

In the fibers with circle-segment cross-sections, the principle of magnetization reversal via domain wall propagation and motion is kept. However, some details differ from the processes simulated for the circular cross-sections.

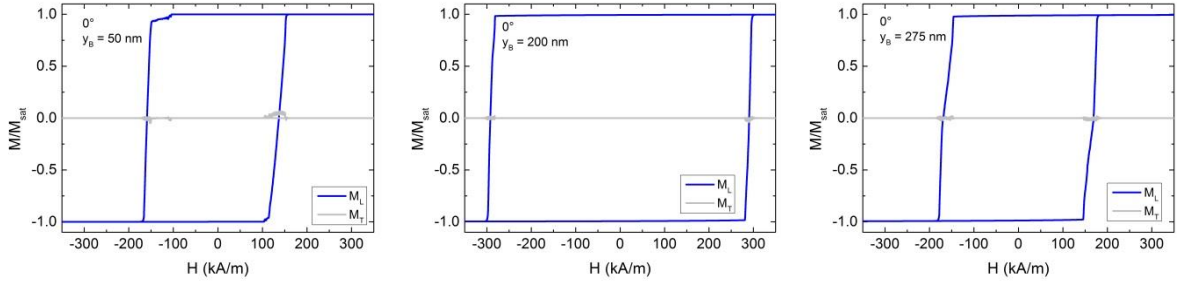

Fig. 15. (Color online) Longitudinal ($M_L$) and transverse ($M_T$) hysteresis loops of fibers with circle-segment cross-section, simulated for an orientation of 0° with respect to the external magnetic field and different bending deflections $y_b$. $M_T$ is nearly constantly zero here.

Fig. 15 depicts the hysteresis loops for a magnetic field orientation of 0° and different values of $y_b$. All graphs seem to stem from easy axis loops. For 50 nm, a slight asymmetry is visible; a feature which can often be recognized in these and other nanostructures.

Figs. 16-18 show snapshots of the magnetization reversal processes for the bending deflections chosen for Fig. 15. Opposite to Figs. 4-6, here no vortex processes are visible – which is understandable due to the modified cross-section, inhibiting the formation of a vortex. Instead, for all values of $y_b$ similar processes take place – which can explain the regular deflection dependence of $H_C$ without jumps in the curves in Fig. 2.

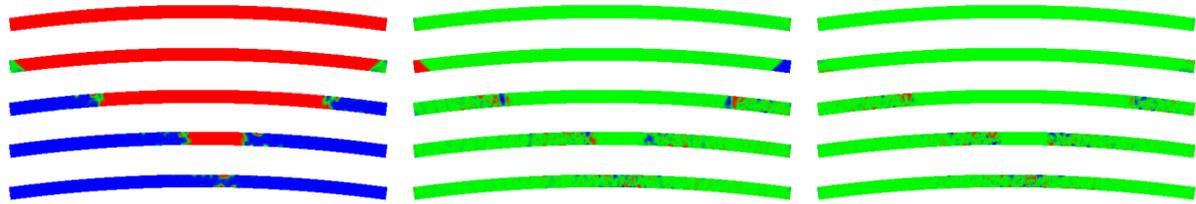

Fig. 16. (Color online) Snapshots of the magnetization for $y_b$ = 50 nm and an angle of 0°, starting from positive saturation (top) to negative saturation (bottom), depicting x, y and z components of the magnetization (from left to right).

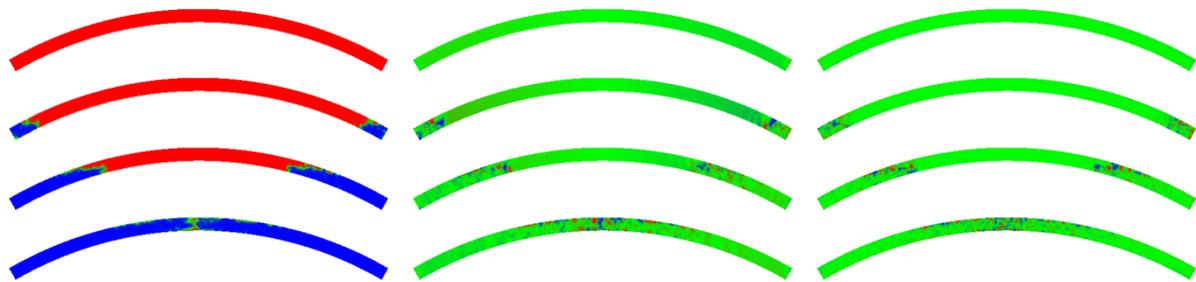

Fig. 17. (Color online) Snapshots of the magnetization for $y_b$ = 200 nm and an angle of 0°, starting from positive saturation (top) to negative saturation (bottom), depicting x, y and z components of the magnetization (from left to right).

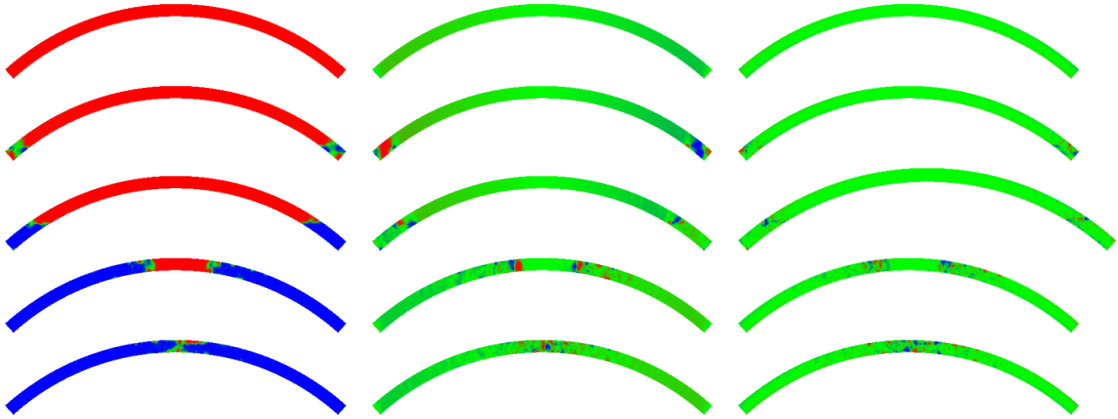

Fig. 18. (Color online) Snapshots of the magnetization for $y_b$ = 275 nm and an angle of 0°, starting from positive saturation (top) to negative saturation (bottom), depicting x, y and z components of the magnetization (from left to right).

Next, Fig. 19 shows the loops for an orientation of 45°. Here, distinct steps in the longitudinal loop are visible for $y_b$ = 350 nm, while an unusual shape of the transverse hysteresis loop can be recognized for $y_b$ = 425 nm.

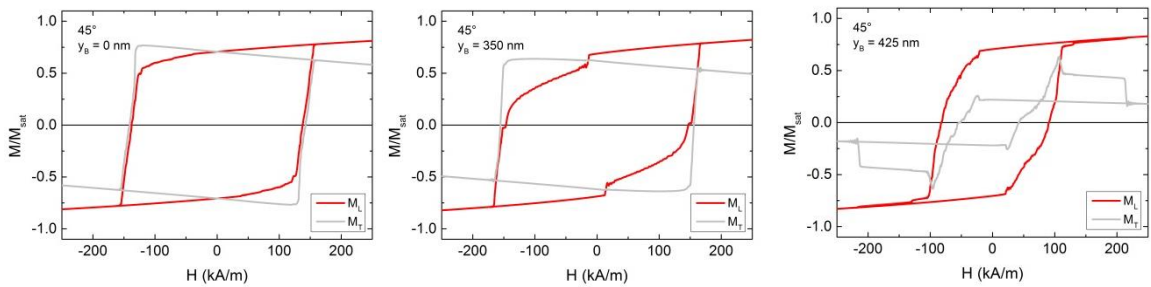

Fig. 19. (Color online) Longitudinal ($M_L$) and transverse ($M_T$) hysteresis loops of fibers with circle-segment cross-section, simulated for an orientation of 45° with respect to the external magnetic field and different bending deflections $y_b$.

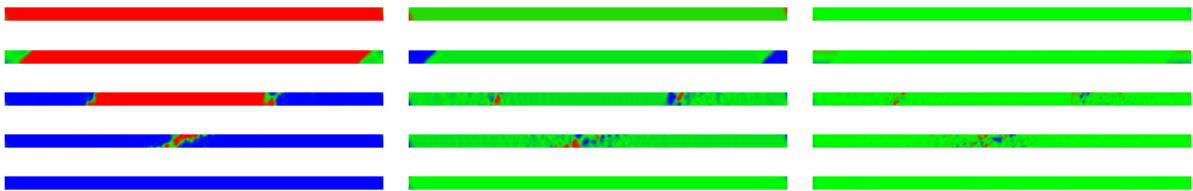

Fig. 20. (Color online) Snapshots of the magnetization for $y_b$ = 0 nm and an angle of 45°, starting from positive saturation (top) to negative saturation (bottom), depicting x, y and z components of the magnetization (from left to right).

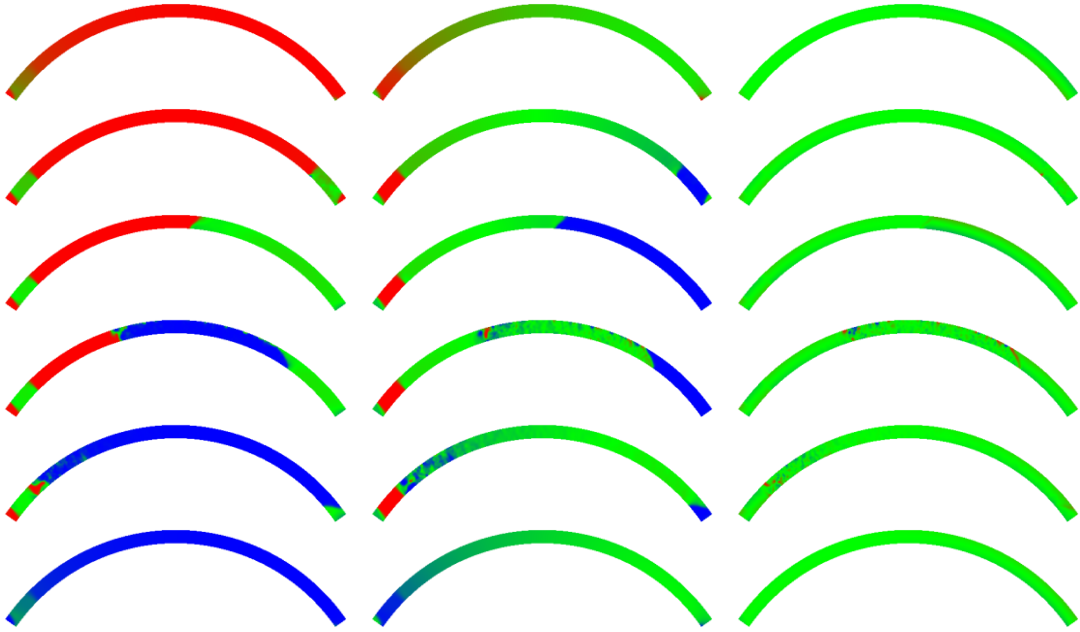

Fig. 21. (Color online) Snapshots of the magnetization for $y_b$ = 350 nm and an angle of 45°, starting from positive saturation (top) to negative saturation (bottom), depicting x, y and z components of the magnetization (from left to right).

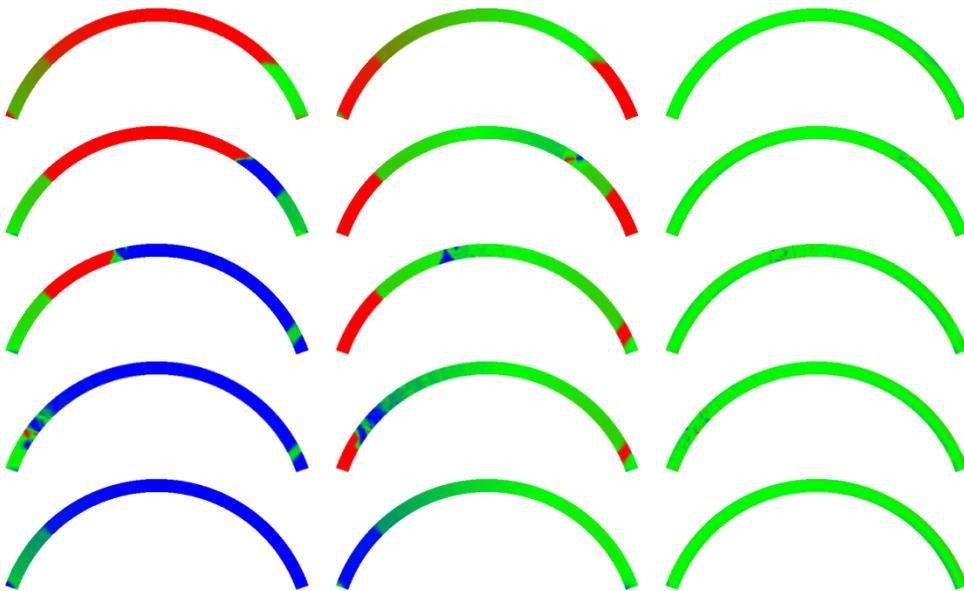

Fig. 22. (Color online) Snapshots of the magnetization for $y_b$ = 425 nm and an angle of 45°, starting from positive saturation (top) to negative saturation (bottom), depicting x, y and z components of the magnetization (from left to right).

Figs. 20-22 depict the corresponding magnetization reversal snapshots. Again, no sign of a vortex formation is visible; magnetization reversal occurs for all deflections by domain wall formation and propagation. For the fibers with $y_b > 0$, again an asymmetry in the magnetization reversal is visible, giving rise to the steps in the hysteresis loops in the field range of the domain wall propagation.

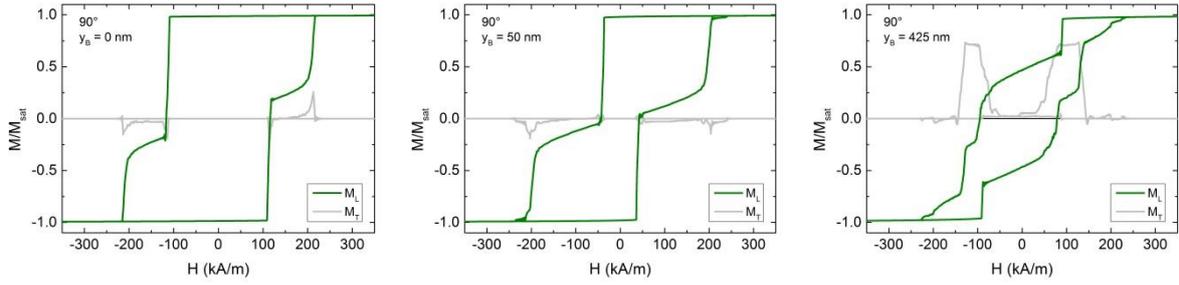

Fig. 23. (Color online) Longitudinal ($M_L$) and transverse ($M_T$) hysteresis loops of fibers with circle-segment cross-section, simulated for an orientation of 90° with respect to the external magnetic field and different bending deflections $y_b$.

In the hysteresis loops measured for a field orientation of 90°, opposite to the same orientation simulated for a circular fiber, steps can be found in all longitudinal hysteresis loops, corresponding to sometimes pronounced transverse peaks.

Figs. 24-26 show the reason for this difference from the magnetization reversal process in the circular fibers: In the circle-segment cross-section simulated here, magnetization is oriented along y (i.e. parallel to the external magnetic field) in the saturated state for all bending deflections. This leads to a two-step reversal process, starting with a two domain walls and the middle domain firstly switching.

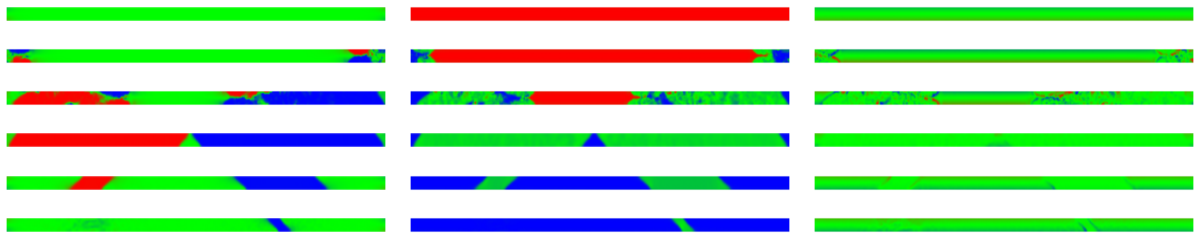

Fig. 24. (Color online) Snapshots of the magnetization for $y_b$ = 0 nm and an angle of 90°, starting from positive saturation (top) to negative saturation (bottom), depicting x, y and z components of the magnetization (from left to right).

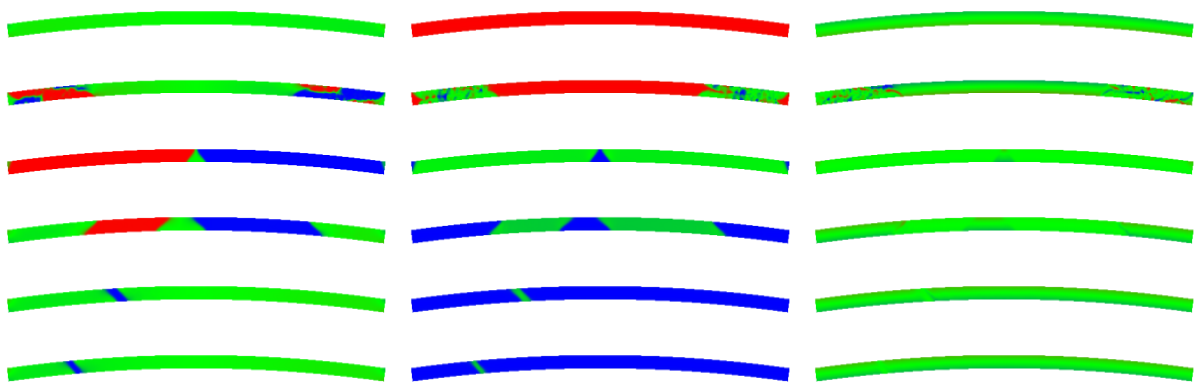

Fig. 25. (Color online) Snapshots of the magnetization for $y_b$ = 50 nm and an angle of 90°, starting from positive saturation (top) to negative saturation (bottom), depicting x, y and z components of the magnetization (from left to right).

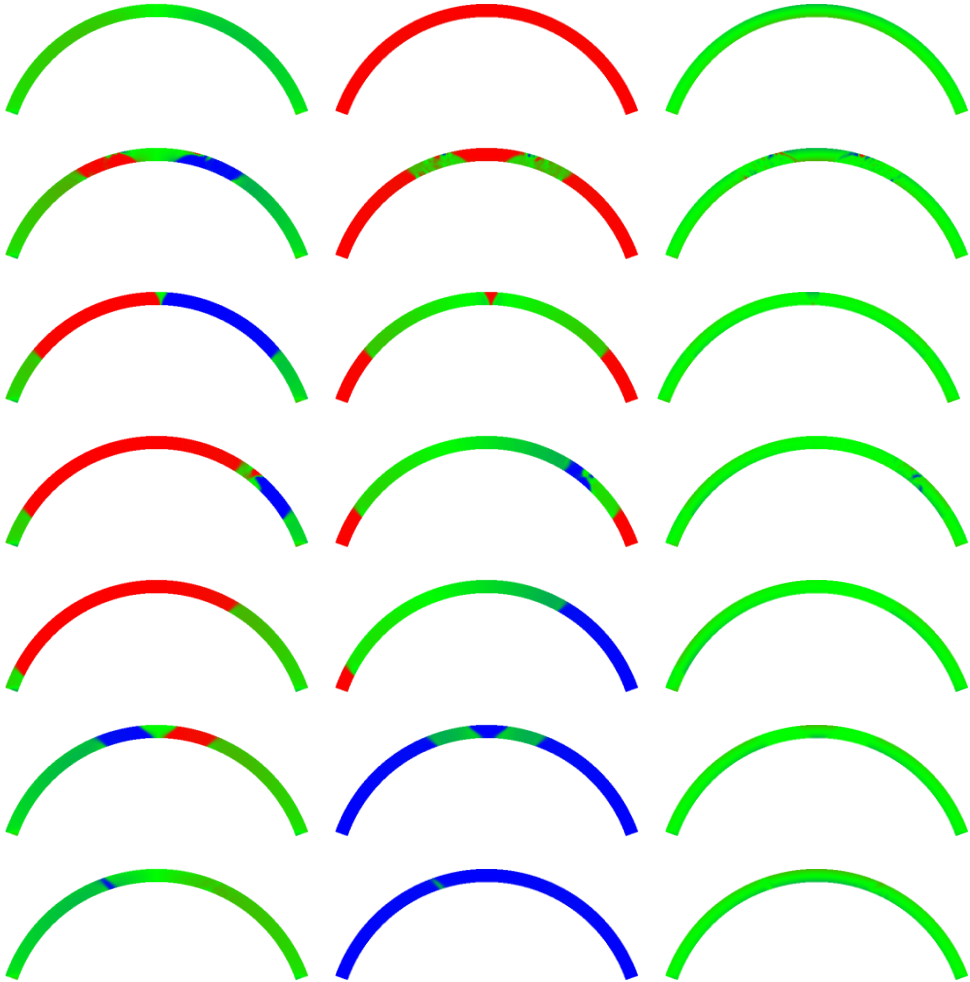

Fig. 26. (Color online) Snapshots of the magnetization for $y_b = 425$ nm and an angle of 90°, starting from positive saturation (top) to negative saturation (bottom), depicting x, y and z components of the magnetization (from left to right).

For the highest $y_b$ value (Fig. 26), an additional step (6$^{th}$ line) is visible which was not recognized before, explaining the additional step in the respective hysteresis loop and the unusual transverse hysteresis loop simulated for $y_b = 425$ nm.

## C. Rectangular cross-section

In the hysteresis loops of 0° sample orientation, a new feature becomes visible for the larges bending deflection. Opposite to all $y_b$ values tested for the circle-segment cross-section, a step becomes visible here, while for smaller deflections, typical easy axis loops can be identified.

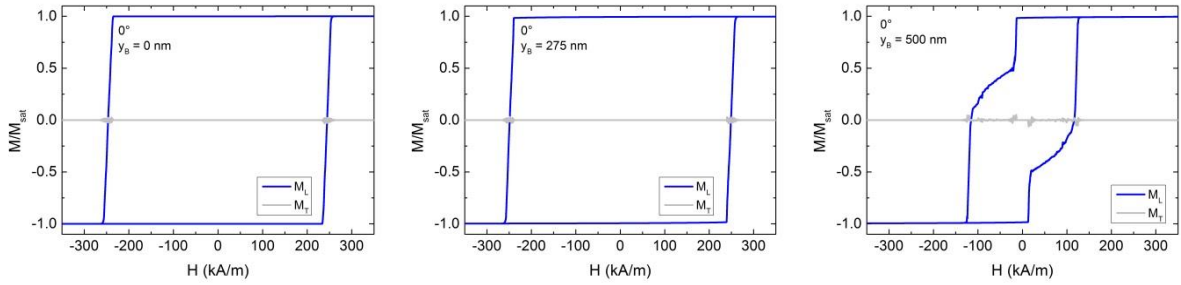

Fig. 27. (Color online) Longitudinal ($M_L$) and transverse ($M_T$) hysteresis loops of fibers with rectangular cross-section, simulated for an orientation of 0° with respect to the external magnetic field and different bending deflections $y_b$. $M_T$ is nearly constantly zero here.

Figs. 28-30 show the corresponding magnetization reversal snapshots. For smaller values of $y_b$ (Figs. 28 and 29), similar processes are visible as for the same orientation simulated for circle-segment cross-sections.

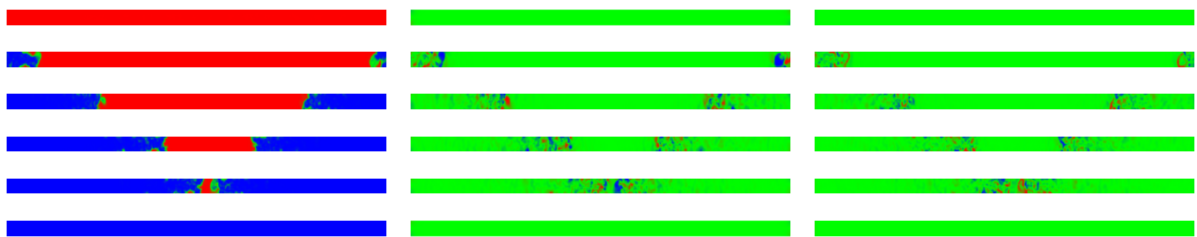

Fig. 28. (Color online) Snapshots of the magnetization for $y_b = 0$ nm and an angle of 0°, starting from positive saturation (top) to negative saturation (bottom), depicting x, y and z components of the magnetization (from left to right).

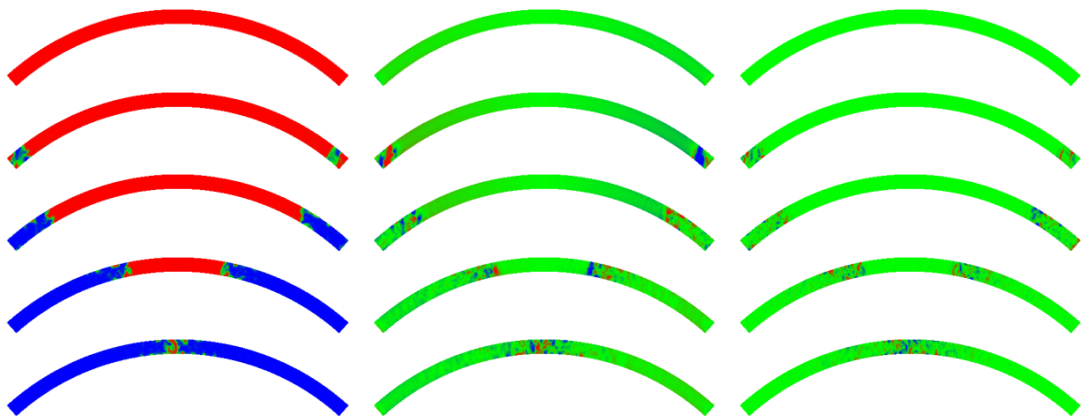

Fig. 29. (Color online) Snapshots of the magnetization for $y_b = 275$ nm and an angle of 0°, starting from positive saturation (top) to negative saturation (bottom), depicting x, y and z components of the magnetization (from left to right).

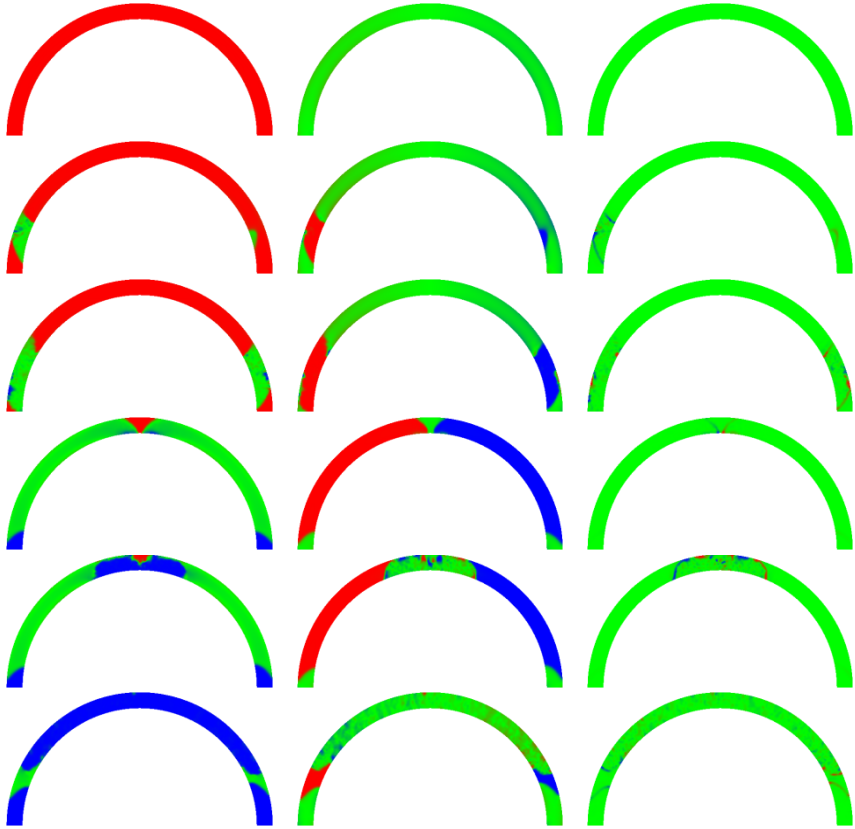

Fig. 30. (Color online) Snapshots of the magnetization for $y_b$ = 500 nm and an angle of 0°, starting from positive saturation (top) to negative saturation (bottom), depicting x, y and z components of the magnetization (from left to right).

For the largest deflection, however, an additional state occurs with the domain wall in the sample middle separating two opposite magnetization orientations along y and in the opposite direction, resulting from the shape anisotropy being partly perpendicular to the magnetic field. This finding explains the small increase of the coercive field for $y_b$ = 500 nm (Fig. 2, rectangular cross-section).

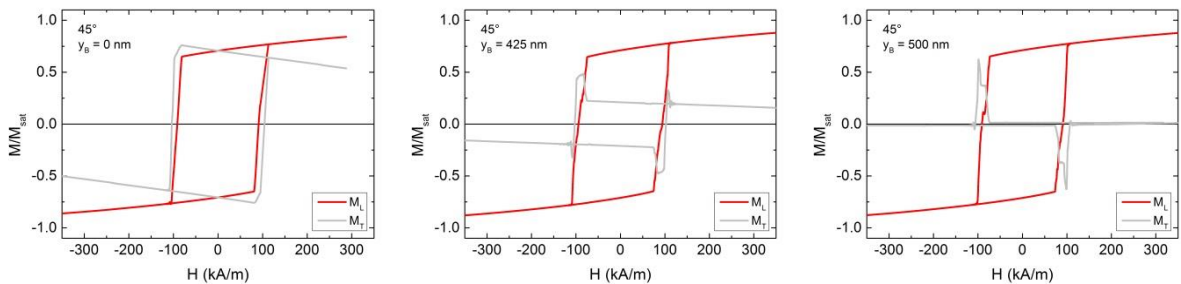

Fig. 31. (Color online) Longitudinal ($M_L$) and transverse ($M_T$) hysteresis loops of fibers with rectangular cross-section, simulated for an orientation of 45° with respect to the external magnetic field and different bending deflections $y_b$.

Next, Fig. 31 shows the hysteresis loops of the 45° orientation, looking quite similar to those simulated for the circular cross-section. Nevertheless, the snapshots of the magnetization reversal processes (Figs. 32-34) show only the expected, slightly asymmetric domain wall formation at both ends of the fibers and the propagation to the inner part of the samples, followed by complete magnetization reversal. No distinct differences are visible between the simulations for varying bending deflections.

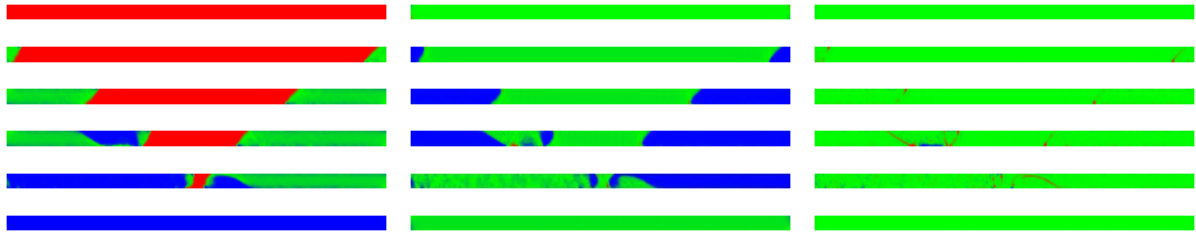

Fig. 32. (Color online) Snapshots of the magnetization for $y_b = 0$ nm and an angle of 45°, starting from positive saturation (top) to negative saturation (bottom), depicting x, y and z components of the magnetization (from left to right).

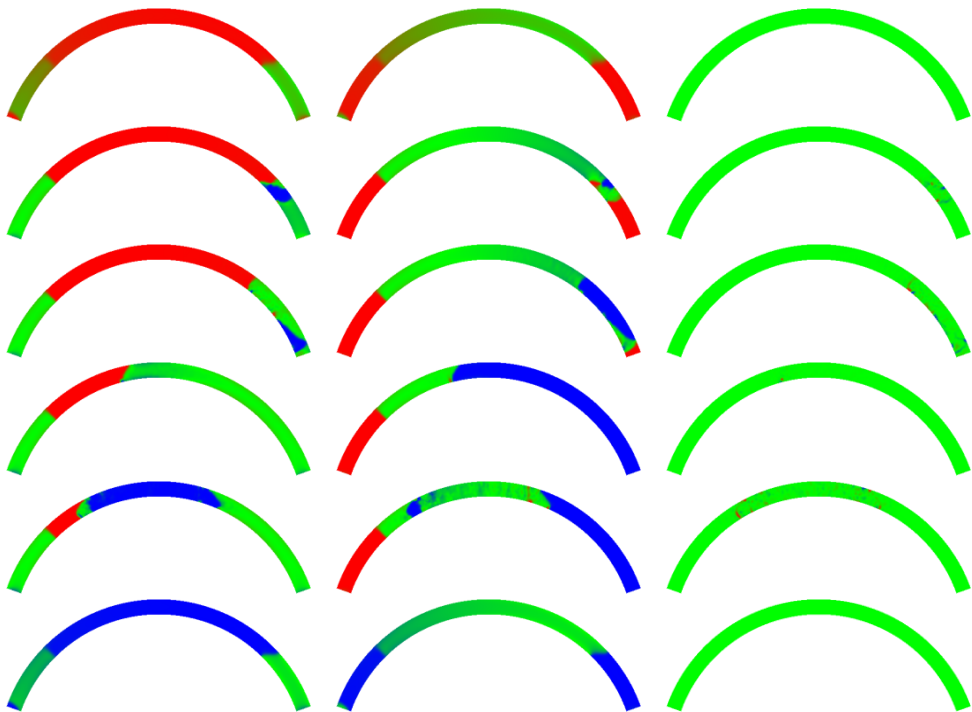

Fig. 33. (Color online) Snapshots of the magnetization for $y_b = 425$ nm and an angle of 45°, starting from positive saturation (top) to negative saturation (bottom), depicting x, y and z components of the magnetization (from left to right).

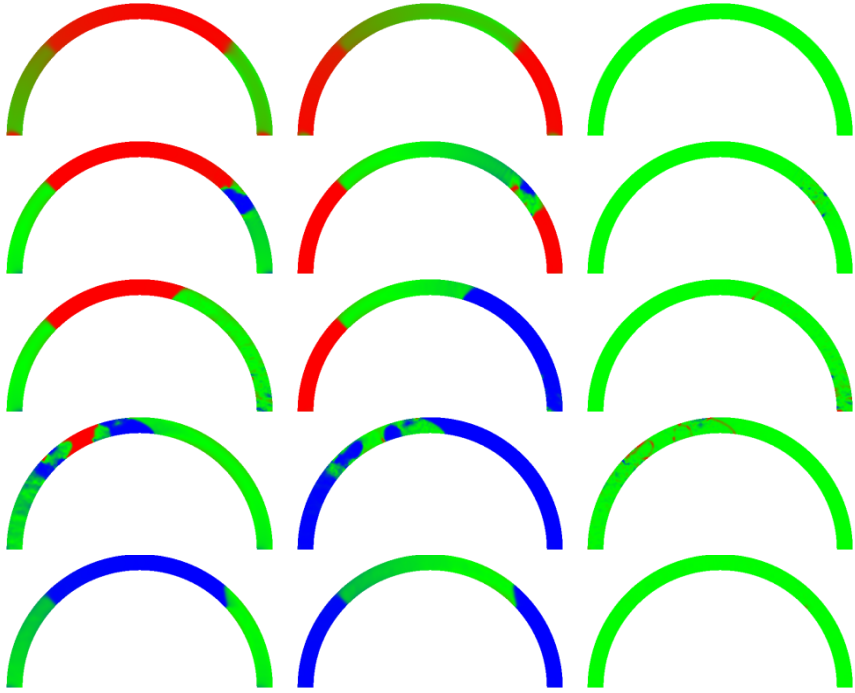

Fig. 34. (Color online) Snapshots of the magnetization for $y_b$ = 500 nm and an angle of 45°, starting from positive saturation (top) to negative saturation (bottom), depicting x, y and z components of the magnetization (from left to right).

Finally, Fig. 35 shows 90° orientation hysteresis loops. Starting with a typical easy axis loop at zero deflection, higher values of $y_b$ result in one or two steps in the loops and corresponding transverse magnetization peaks.

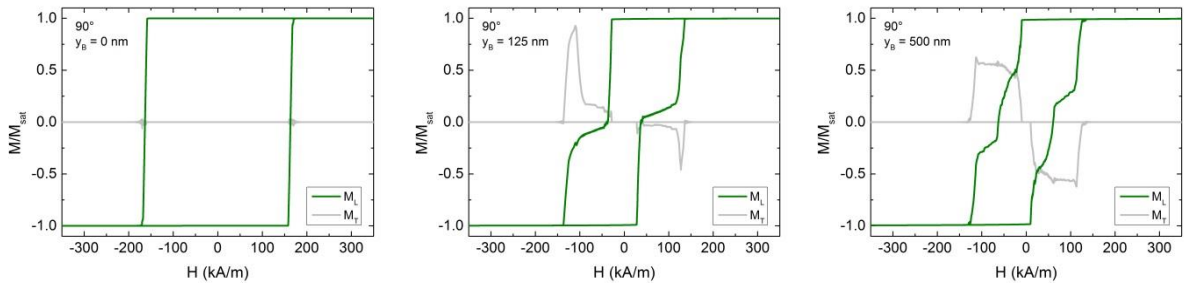

Fig. 35. (Color online) Longitudinal ($M_L$) and transverse ($M_T$) hysteresis loops of fibers with rectangular cross-section, simulated for an orientation of 90° with respect to the external magnetic field and different bending deflections $y_b$.

Correspondingly, Fig. 36 shows the typical 90° snapshots, now interestingly again with vortex domain walls along the x direction which are also weakly visible in the z direction. Apparently, the rectangular cross-section seems to be located between circular and circle-segment cross-section, showing a mixture between the magnetization reversal processes typical for the other two situations.

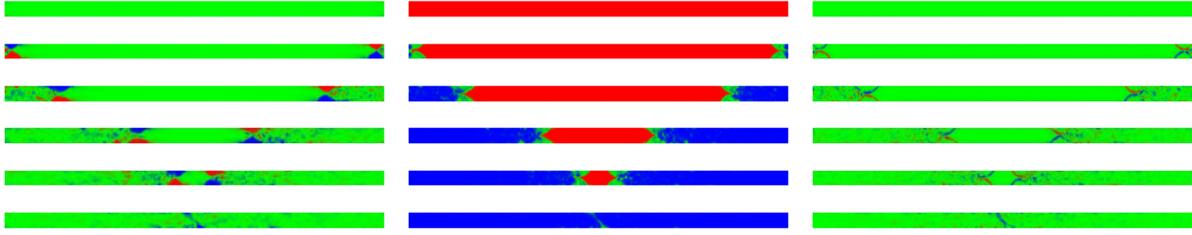

Fig. 36. (Color online) Snapshots of the magnetization for $y_b = 0$ nm and an angle of 90°, starting from positive saturation (top) to negative saturation (bottom), depicting x, y and z components of the magnetization (from left to right).

In Figs. 37 and 38, reversal processes for higher bending deflections are depicted, showing again one- and two-domain wall processes, similar to the situation found for 90° in the circle-segment cross-section fibers.

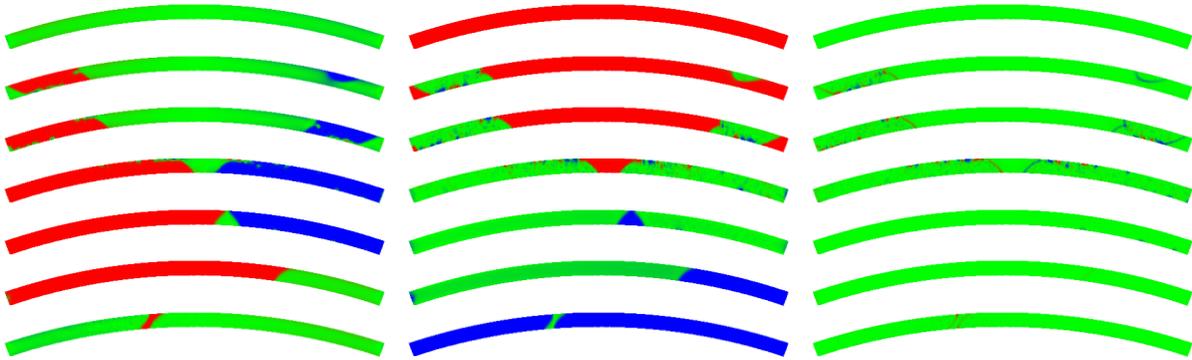

Fig. 37. (Color online) Snapshots of the magnetization for $y_b = 125$ nm and an angle of 90°, starting from positive saturation (top) to negative saturation (bottom), depicting x, y and z components of the magnetization (from left to right).

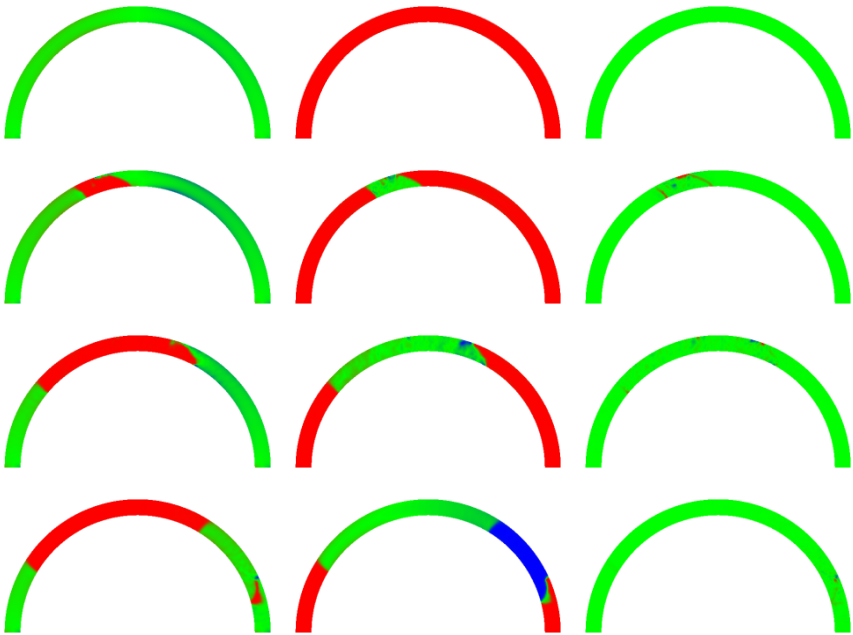

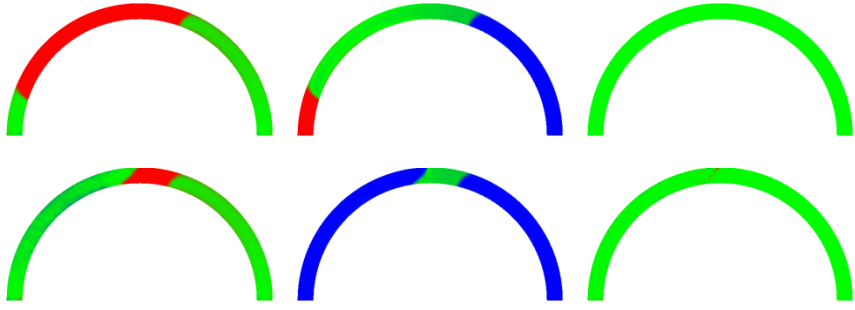

Fig. 38. (Color online) Snapshots of the magnetization for $y_b$ = 500 nm and an angle of 90°, starting from positive saturation (top) to negative saturation (bottom), depicting x, y and z components of the magnetization (from left to right).

As the above described simulations have shown, magnetization reversal processes in magnetic nanofibers depend strongly on their cross-sections and the bending deflections as well as the orientation of the fibers with respect to the external magnetic field. Thinking about utilizing such nanofibers in an arbitrary arrangement for data storage and/or calculation is thus a large challenge, necessitating more detailed knowledge about the magnetization reversal processes under different conditions – especially for such fibers being embedded in a large network of ferromagnetic and potentially conductive fibers.

## IV. Conclusion

Using fiber networks may serve as a possible base of a new hardware for cognitive (bio-inspired) computing without the strict separation between memory and processor in the classical von Neumann architecture, similar to the human brain. Biologically inspired algorithms and software, however, have only been implemented yet on common computer hardware, restricting the possibilities of such new neuromorphic approaches.

This is why we have investigated nanofibers of different cross-sections and bending deflections under different angles with respect to an external magnetic field. While domain wall processes are responsible for the magnetization reversal in all cases, the details differ strongly. Different types of domain walls, including vortex domain walls, were found; different numbers of domain walls can co-exist in various fiber geometries. Since domain walls can be used to store and manipulate data, the principal possibility to use such fiber-based geometries for date storage and processing is given.

As a next step, fiber networks will be investigated, using geometrically arranged as well as randomly oriented fibers, to examine further the possibilities to use such nanofiber networks for data storage and manipulation.

**Acknowledgment**
The work was partially supported by the SUT Rector Grant 14/990/RGJ17/0075-02 (T. B.).